\documentclass[universe,review,accept,pdftex,oneauthor]{Definitions/mdpi} 
\firstpage{1} 
\pubvolume{1}
\issuenum{1}
\articlenumber{0}
\pubyear{2026}
\copyrightyear{2026}
\externaleditor{ } % More than 1 editor, please add `` and '' before the last editor name Firstname Lastname
\datereceived{ } 
\daterevised{ } % Comment out if no revised date
\dateaccepted{ } 
\datepublished{ } 

\usepackage{changepage}  % For adjustwidth

\usepackage{luatex85}

\usepackage{tikz}

\newcommand{\ha}  {H$\alpha$}
\newcommand{\ew}  {EW(H$\alpha$)}

\def\msun {{\mathrm{M}_\odot}}
\def\sun {{_{\odot}}}
\def\rsun {{R\sun}}
\def\simless{\mathbin{\lower 3pt\hbox
     {$\rlap{\raise 5pt\hbox{$\char'074$}}\mathchar"7218$}}}   %< or of order
\def\simmore{\mathbin{\lower 3pt\hbox
     {$\rlap{\raise 5pt\hbox{$\char'076$}}\mathchar"7218$}}}   %> or of order

\Title{Be/X-Ray Binaries: Phenomenology, Variability, and Accretion~Dynamics}

\Author{Pablo Reig $^{1,2}$\orcidA}

\AuthorNames{Pablo Reig}

\address{%
$^{1}$ \quad Institute of Astrophysics, Foundation for Research and Technology-Hellas, 
   	71110 	Heraklion, Greece; pau@physics.uoc.gr
\\
$^{2}$ \quad Physics Department, University of Crete, 71003 Heraklion, Greece}

\abstract{
Be/X-ray binaries constitute the largest and most diverse subgroup of neutron
star high-mass X-ray binaries. These systems feature a rapidly rotating Be star
surrounded by a circumstellar decretion disk that serves as the primary
reservoir of accreted matter onto a strongly magnetized neutron star.   While a
few Be/X-ray binaries remain persistently active, the majority manifest as hard
X-ray transient sources, becoming detectable only during X-ray outbursts. This
review synthesizes current understanding of their rich phenomenology across
optical and X-ray wavelengths, focusing on variability ocurring over timescales
that span from seconds to years. }

\keyword{X-rays binaries; neutron stars;  emission line; Be stars}

\begin{document}

\section{Classification of X-Ray~Binaries}

X-ray binaries consist of a compact object orbiting around a ``normal'' star.
They are ``close'' binary systems because there exists a transfer of mass from the
optical component to the compact object. By~``normal'' star, it is meant that
nuclear burning is still occurring in its interior. The~compact object can be a
black hole, a~neutron star, or~a white dwarf, and~we refer to them as black-hole
binaries (BHBs), neutron star binaries (NSBs), or~cataclysmic variables (CVs),
respectively. The~spectral type of the optical star defines the system as either
a low-mass X-ray binary (LMXB) or a high-mass X-ray binary (HMXB). In~LMXBs, the~spectral type of the optical companion is typically later than {\em A}, while in
HMXBs it is an early-type {\em B} or late-type {\em O} star. In~HMXBs, the~mass of
the companion is typically larger than $\sim$8 $\msun$.  Our Galaxy hosts a few
BHBs (only one confirmed: Cyg X--1, plus candidates like Cyg X--3 and SS 433)
and about half of the NSBs are found in HMXBs. LMXBs, with~the mass of the companion below $\sim$2
$\msun$ include the majority of BHBs and half of the NSBs. CVs contain low-mass
companions and are normally considered a different class (the term ``X-ray
binaries'' is normally reserved for binaries with neutron stars or black holes). 

Neutron star high-mass X-ray binaries (NS-HMXBs) are accretion-powered systems
where a neutron star accretes matter from a massive, early-type companion. They
rank among the brightest and most variable X-ray sources in the Milky Way and
nearby galaxies and serve as key laboratories for high-energy astrophysics by
combining several extreme and interacting physical processes in a single
observable system. Studies of NS-HMXBs cover a broad range of topics,
from~accretion physics to stellar evolution and galactic structure, making them
central to both high-energy and stellar astrophysics. The~vast majority of
NS-HMXBs are X-ray binary transient~systems.  

NS-HMXBs are classified according to the luminosity class of the donor star into
supergiant X-ray binaries (SGXBs) and Be/X-ray binaries (BeXBs). In~supergiant
systems, the~neutron star orbits an evolved O or B supergiant whose powerful
stellar wind provides the material for accretion, typically through wind
capture (low-luminosity SGXBs), although~in a few cases, Roche-lobe overflow leads to the formation of
an accretion disk and enhanced X-ray emission (high-luminosity SGXBs). This class also includes the fast and
highly variable systems known as supergiant fast X-ray transients.  This article
focuses on the observational properties of Be/X-ray binaries (BeXBs), the~most
numerous subgroup of NS-HMXBs. In~BeXBs, the~optical companion is a Be star,
characterized by emission lines (most notably the H$\alpha$ line), an infrared
excess (increased flux at longer wavelengths), and~a small degree of linear
polarization (typically less than 2\%).  These features arise from a
circumstellar disk, also  known as decretion disk, around its equatorial plane.
The~disk forms from material ejected from the stellar photosphere and evolves
over years, undergoing formation, growth, and~dissipation. When the disk
dissipates, emission lines revert to absorption, the~infrared excess disappears,
and polarization is no longer detected. Since the disk acts as the reservoir of
matter for accretion, it fundamentally drives the observed X-ray variability.
In~other words, the~accretion process is governed by the interaction between the
neutron star and this decretion~disk.

\begin{figure}[H]
%\centering
\includegraphics[width=.7\linewidth]{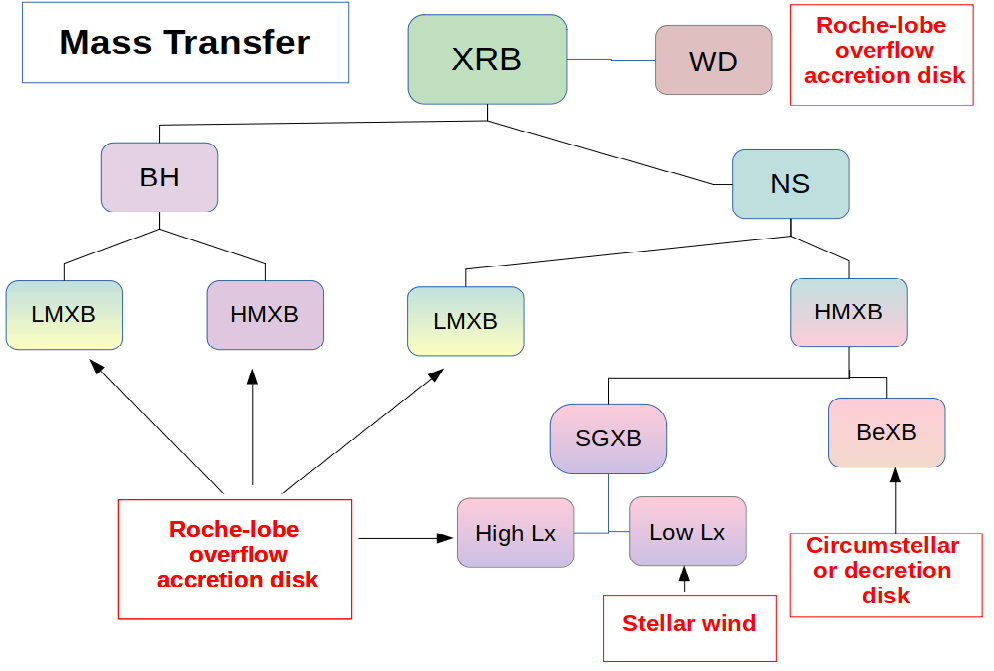} 
\caption{Taxonomy
 of X-ray~binaries\label{class}}
\end{figure}

Figure~\ref{class} shows a tree diagram that illustrates the different types of
X-ray binaries and the mass transfer mechanism. 
Table~\ref{bexb-list} lists the Galactic BeXBs with confirmed optical
counterparts. Only systems for which at least one optical spectrum is available
are included. The~\ew\ is the maximum value reported in the literature and is
normally related to bright X-ray emision. However, note that for some systems,
e.g, MAXI\,J0903--531, RX\,J1037.5--5647 only one or very few measurements are
available and correspond to relatively quiescence X-ray~states.

%--------------------------------------------------------------
\unskip   
%--------------------------------------------------------------

%-------------------------------------------------------------
\begin{table}
%\small
%\footnotesize
\scriptsize
\caption{List of BeXBs with confirmed optical counterparts. 
\label{bexb-list}}
%\centering          

\begin{adjustwidth}{-\extralength}{0cm}
\centering %% If there is a figure in wide page, please release command \centering, for Table, ``\textwidth" should be ``\fulllength"
%\begin{tabular}{l@{~~}l@{~~}c@{~~}c@{~~}c@{~~}c@{~~}c@{~~}c@{~~}}
\begin{tabularx} {\fulllength} {llCCCCcc}
\toprule%\hline
\textbf{Source}          	&\textbf{Spectral}   &\boldmath{$V$}\textbf{-Band}	&\textbf{P}\boldmath{$_{\rm spin}$} &\textbf{P}\boldmath{$_{\rm orb}$} &\boldmath{$e$}   & \textbf{Distance (kpc)}  & \boldmath{$-$}\textbf{\ew}		\\
\textbf{Name}            	&\textbf{Type}       &\textbf{(mag)}	&\textbf{(s) }   	& \textbf{(days)}      &       &\textbf{Gaia DR3}        &\textbf{(\AA)}	\\
\midrule
4U\,0115+63     	&B0.2Ve     &15.4	&3.6    &24.3	&0.34   &$5.7^{+0.5}_{-0.4}$         &$17\pm2$  	\\            \midrule
IGR\,J01363+6610	&B1IV-Ve    &13.3	&-- 	&--	&--     &$5.6\pm0.4$	             &$72\pm3$  	\\            \midrule
RX\,J0146.9+6121	&B1Ve	    &11.3	&1400   &330~$^{\dag\dag}$	&--     &$2.8\pm0.2$ 		     &$17\pm1$  	\\         \midrule
IGR\,J01583+6713	&B2IVe	    &14.4	&--  	&--	&--     &$5.6^{+0.6}_{-0.4}$         &$75\pm3$  	\\            \midrule
RX\,J0240.4+6112	&B0Ve       &10.8	&--     &26.5	&0.54   &$2.50\pm0.07$	             &$17\pm1$  	 \\            \midrule
Swift\,J0243.6+6124	&O9.5Ve     &12.8	&9.9    &27.6	&0.10   &$5.2^{+0.3}_{-0.2}$         &$11\pm1$  	 \\

%\bottomrule
%\end{tabularx}%} % if “}” necessary
%\end{adjustwidth}
%\end{table}

%\begin{table}[H]\ContinuedFloat
%%\tablesize{\small}
%%\footnotesize
%\scriptsize
%\caption{\textit{Cont.}}
%%\begin{adjustwidth}{-\extralength}{0cm}
%%\centering
%\begin{adjustwidth}{-\extralength}{0cm}
%\centering %% If there is a figure in wide page, please release command \centering, for Table, ``\textwidth" should be ``\fulllength"
%%\begin{tabular}{l@{~~}l@{~~}c@{~~}c@{~~}c@{~~}c@{~~}c@{~~}c@{~~}}
%\begin{tabularx} {\fulllength} {llCCCCcc}
%\toprule%\hline
%\textbf{Source}          	&\textbf{Spectral}   &\boldmath{$V$}\textbf{-Band}	&\textbf{P}\boldmath{$_{\rm spin}$} &\textbf{P}\boldmath{$_{\rm orb}$} &\boldmath{$e$}   & \textbf{Distance (kpc)}  & %\boldmath{$-$}\textbf{\ew}		\\
%\textbf{Name}            	&\textbf{Type}       &\textbf{(mag)}	&\textbf{(s) }   	& \textbf{(days)}      &       &\textbf{Gaia DR3}        &\textbf{(\AA)}	\\

\midrule
V\,0332+53      	&O8--9Ve    &15.4	&4.4    &33.8	&0.37   &$5.9\pm0.4$   		     &$8.5\pm0.5$	 \\            \midrule
X\,Per          	&B0Ve       &6.1	&838    &250	&0.11   &$0.60^{+0.02}_{-0.01}$      &$40\pm2$  	 \\            \midrule
RX\,J0440.9+4431	&B0.2Ve     &10.7	&202.5  &150	&--     &$2.44^{+0.08}_{-0.09}$      &$13.5\pm0.5$	  \\            \midrule
1A\,0535+262    	&O9.7IIIe   &9.2	&105    &111    &0.47   &$1.77\pm0.06$		     &$24\pm1$  	 \\            \midrule
IGR\,J06074+2205	&B0.5Ve     &12.2	&373.2  &80~$^{\dag}$     &--     &$6.0^{+0.9}_{-0.6}$         &$12\pm1$  	  \\            \midrule
MXB\,0656--072  	&O9.5Ve     &12.3	&160.4  &101.2  &0.4~$^{\dag\dag}$   &$5.7\pm0.5$	             &$25\pm2$  	 \\            \midrule
MAXI\,J0655--013	&O9.5-B0Ve  &11.7	&1130	&27.9	&0.48	&$3.7^{+0.9}_{-0.6}$	     &$33\pm3$  	 \\            \midrule
4U\,0728--25		&O8--9Ve    &11.3	&103.3	&34.5	&--	&$6.7\pm2.3$		     &$9\pm1$		\\            \midrule
2SXPS\,J075542.5--293353&B0Ve	    &10.2	&307.8	&59.7	&0.06	&$3.3\pm0.2$		     &$23\pm2$  	\\            \midrule
RX\,J0812.4--3114	&B0.2IVe    &12.7	&31.9   &80	&--     &$6.7^{+0.5}_{-0.4}$         &$20\pm1$  	 \\            \midrule
GS\,0834--430		&B0--2III-Ve&20.4	&12.3	&105.8	&0.15	&$5.7^{+2.4}_{-1.6}$	     &$32\pm2$  	\\            \midrule
MAXI\,J0903--531	&B1.5--2III-Ve&13.3	&14.05	&57.0	&--	&$9.2^{+3.3}_{-3.8}$	     &$4.5\pm0.5$	\\            \midrule
GRO\,J1008--57   	&B1Ve       &15.3 	&93.5   &249.5	&0.68   &$3.5\pm0.2$	             &$29\pm1$  	 \\            \midrule
RX\,J1037.5--5647	&B0III-Ve   &11.5	&860    &61 	&--     &$5.1\pm0.4$		     &$1.5\pm0.5$	\\            \midrule
1A\,1118--616   	&O9.5IV-Ve  &12.1	&406.5 	&24	&$<0.1$ &$2.90^{+0.09}_{-0.08}$      &$90\pm10$ 	\\            \midrule
IGR\,J11435--6109	&B0.5Ve	    &16.4	&161.8	&52.4	&--	&$7.7^{+1.7}_{-1.9}$	     &$26\pm2$  	\\            \midrule
4U\,1145--61    	&B0.2IIIe   &9.0	&292    &186.5	&$>0.5$ &$2.06^{+0.08}_{-0.09}$      &$45\pm3$  	\\            \midrule
4U\,1258--61(GX\,304--1)&B2III-Ve   &14.4	&272    &132.5	&0.5	&$1.90\pm0.05$	             &$39\pm2$  	\\            \midrule
2S\,1417--624		&B1Ve	    &16.9	&17.7	&42.12	&0.45	&$6.4^{+3.6}_{-2.1}$	     &$12\pm2$  	\\            \midrule
Swift\,J1626.6-5156	&B0Ve	    &15.5	&15.4	&132.9	&0.08	&$7.9^{+4.1}_{-2.0}$	     &$40\pm2$  	\\            \midrule
XTE\,J1946+274		&B0--1V-IVe  &16.6	&15.8	&172	&0.25	&$12.1^{+2.6}_{-2.2}$	     &$52\pm4$  	\\            \midrule
KS\,1947+300    	&B0Ve       &14.5	&18.7   &40.4	&0.03   &$14^{+3}_{-2}$		     &$19\pm1$  	\\            \midrule
Swift\,J2000.6+3210	&B0--2Ve     &16.2	&890   	&-- 	&--     &$8.4^{+1.8}_{-1.2}$         &$15\pm2$  	\\            \midrule
EXO\,2030+375		&B0Ve	    &19.5	&42	&46.02	&0.41	&$2.4^{+0.5}_{-0.4}$	     &$21\pm2$  	\\            \midrule
GRO\,J2058+42   	&O9.5-B0IV-Ve&14.9	&192	&55	&--	&$8.9^{+0.7}_{-0.8}$	     &$15\pm1$  	\\            \midrule
SAX\,J2103.5+4545	&B0Ve       &13.9	&358	&12.7	&0.40   &$6.2^{+0.4}_{-0.5}$	     &$5.5\pm0.5$	\\            \midrule
IGR\,J21343+4738	&B1IVe      &14.1	&320	&--	&--	&$8.5^{+1.1}_{-0.8}$	     &$15\pm5$  	\\            \midrule
Cep\,X--4		&B1--2Ve     &14.3	&66.3   &--	&--     &$7.2^{+0.68}_{-0.6}$        &$55\pm1$  	\\            \midrule
4U\,2206+54     	&O9.5Ve	    &9.8	&5550   &19.2	&0.15   &$3.2^{+0.2}_{-0.1}$         &$5.0\pm0.5$	\\            \midrule
SAX\,J2239.3+6116	&B0Ve       &14.4	&1247   &263	&--     &$7.3^{+0.7}_{-0.5}$         &$19\pm2$  	\\            \midrule
IGR\,J22534+6243	&B1Ve	    &15.3	&46.67	&--	&--	&$8.9^{+1.4}_{-0.9}$	     &$39\pm1$  	\\
\midrule%\hline
\end{tabularx}
\end{adjustwidth}
\noindent{\footnotesize Sources: \cite{bailer-jones21,neumann23,fortin23,kim23}, Simbad data base, and~author's own search;
\textsuperscript{$\dag$} $P_{\rm  orb}=80/n$, where $n=2,3,\dots $; 
\textsuperscript{$\dag\dag$} Tentative.
}
\end{table}
\unskip
%-------------------------------------------------------------            

\section{A Historical~View}

Over the past six decades, HMXBs have become central to astrophysics, serving as
laboratories for accretion physics, stellar winds, magnetic fields, binary
evolution, and~the formation of compact objects. Key milestones in the history
of HMXBs include~{\cite{reig26b}}:

\begin{itemize}

    \item {\em Early discoveries: birth of X-ray astronomy (1960s)}. The~field
began in 1962 with the discovery of the first extrasolar X-ray source, Sco X-1~\cite{giacconi62}. A~few years later, Centaurus X-3 (Cen X-3) was discovered,
first as an X-ray source~\cite{chodil67} and then as an X-ray pulsar~\cite{giacconi71}. Early observations were limited by poor angular and spectral
resolution, and~the physical origin of X-ray emission was debated. Accretion
onto compact objects emerged early as a viable mechanism, particularly within
close binary systems, where mass transfer could efficiently power luminous X-ray
emission~\cite{shklovsky67}.

    \item {\em Consolidation of the binary model (1970s)}. During~this decade, the~binary nature of Galactic X-ray sources became widely accepted. Systems were
classified based on the nature of the compact object (white dwarf, neutron
star, or~black hole) and the mass of the donor star, leading to the distinction
between low-mass and high-mass X-ray binaries~\cite{burbidge72,gursky75}. Among~HMXBs, two main subclasses were recognized: systems with supergiant donors
and those with Be-type companions~\cite{maraschi76}. A~major breakthrough was the
discovery of coherent X-ray pulsations and cyclotron resonance scattering
features~\cite{rappaport77, trumper78}, establishing that most HMXBs host
strongly magnetized neutron stars with field strengths of $(1-10) \times 10^{12}$
G.

    \item {\em Mass transfer and phenomenology (1980s)}. By~the 1980s, SGXBs
were understood as persistent sources powered mainly by wind accretion from
luminous supergiants, whereas BeXBs were recognized as transient systems
associated with circumstellar decretion disks around Be stars
~\cite{rappaport82,white89}. Two types of BeXB X-ray outbursts were defined:
periodic, orbitally modulated Type I outbursts near periastron, and~rarer,
brighter, irregular Type II outbursts linked to large-scale disk disturbances~\cite{stella86}. The~realization that distinct HMXB subclasses correspond to
different mass-transfer mechanisms culminated in the Corbet diagram, which
revealed a correlation between neutron-star spin period and orbital period and
provided a unifying framework for HMXB taxonomy~\cite{corbet86}.

    \item {\em Expansion of the BeXB picture (1990s)}. Improved X-ray
sensitivity led to the discovery of persistent, low-luminosity BeXBs~\cite{reig99}. These systems differ from classical transient BeXBs by exhibiting long
spin periods, weak variability, low accretion rates, and~likely wide, nearly
circular orbits, indicating steady accretion from the tenuous outer regions of Be-star
disks rather than episodic disk-neutron star~interactions.

    \item {\em New subclasses and complexity (2000s)}. The~launch of the
INTEGRAL mission marked a turning point, by~revealing previously hidden
populations of SGXBs and introducing new subclasses. Supergiant Fast X-ray
Transients were identified as systems with supergiant donors that
exhibit brief, intense X-ray flares separated by long periods of quiescence~\cite{negueruela06}, challenging the earlier assumption that all SGXBs are
persistent emitters. This decade also saw the recognition of $\gamma$ Cas
analogues---Be stars with anomalously hard, thermal X-ray emission at relatively
low luminosities and lacking detected pulsations, suggesting fundamentally
different emission mechanisms, possibly involving magnetic star-disk
interactions or accretion in unusual regimes~\cite{lopes06}.

    \item {\em Recent developments (2010s)}. The~diversity of HMXBs expanded
further with the identification of $\gamma$-ray binaries, systems emitting
primarily above MeV energies, likely powered by the interaction between a pulsar
wind and the massive companion winds~\cite{dubus13}. Another major discovery was
the first Be-star binary hosting a black hole, enriching BeXB phenomenology
though its nature remains under debate~\cite{casares14,janssens23}.

\end{itemize}

%---------------------------------------------------------------------------

%---------------------------------------------------------------------------

\section{Variability~Timescales}

Variability in a system generally means that some quantity or observable changes
over time. The~timescales of this variability help identify the physical
processes involved and their origins. Variability can be periodic,
quasi-periodic, or~aperiodic. Be/X-ray binaries exhibit variability
across all timescales, from~seconds to years, and~across many different bands of
the electromagnetic~spectrum.

In the X-ray band, variability originates from accretion. Timescales related to
accretion range from fractions of a second to hours, appearing as pulsations,
quasi-periodic oscillations, and~broadband noise in power spectra~\cite{revnivtsev09,mushtukov19,yang26}. Longer timescales (weeks to months)
show up as orbitally modulated Type I X-ray outbursts. The~slowest variability,
on scales of years, is linked to mass transfer between the Be star and neutron
star, manifesting as unpredictable giant or Type II X-ray~outbursts.

In the optical band, short-term variations (hours) are connected to the Be
star's atmosphere and its rapid rotation. Intermediate timescales (weeks to
months) relate to processes in well-developed circumstellar disks, such as the
precession of density perturbations. The~longest timescales (years) correspond
to the formation, growth, and~dissipation of these~disks.

The physical information extracted from variability studies naturally depends on
the observable under investigation. For~long-term monitoring, easy-to-measure
observables are preferred: the \ha\ emission line in optical spectra is
particularly powerful for diagnosing the size, density, and~structure of Be star
decretion disks because it forms over much of the disk and is optically thick.
Its strength and shape (e.g., equivalent width, V/R ratios, peak separation)
reveal disk density, radial extent, and~velocity fields~\cite{quirrenbach97,hummel00,tycner05,grundstrom06}. It is also less affected by
interstellar extinction than some helium or metal lines. From~photometry, magnitudes and colors are easily obtained. In~polarimetry, the~polarization angle and the degree of polarization, can be directly
computed from aperture~photometry.

The variability in BeXBs arises from three main sites: the Be star, the~circumstellar disk, and~the neutron star. The~following review discusses key
physical processes linked to different variability timescales according to their
origin. Table~\ref{var} summarizes these timescales, their physical origins, and~the
processes they~probe.

\begin{table}[H]
\scriptsize
\caption{Typical variability time scales in~BeXBs.\label{var}}
%MDPI: This is a reminder: figures and tables should generally be placed immediately after the paragraph in which they are first mentioned. Please note that further adjustments to the position or size of figures and tables may be made during the production stage to minimize white space. We have revised their positions/sizes accordingly. Please confirm.
%\begin{adjustwidth}{-0.5cm}{0cm}  % Adjust -1cm to your needs
%\begin{tabularx}{\linewidth}{>{\centering\arraybackslash}X
%                             >{\centering\arraybackslash}X
%                             >{\centering\arraybackslash}X
%                             >{\centering\arraybackslash}X}

 \begin{tabularx} {\textwidth} {CccC}                           
                             
\toprule
\multirow{2}{*}{\textbf{Time Scale}} & \multirow{2}{*}{\textbf{Physical Process}} & \textbf{Observational} &  \multirow{2}{*}{\textbf{Origin}} \\
                                     &                                            & \textbf{Evidence}       & \\
\midrule

\multirow{2}{*}{Seconds} & \multirow{2}{*}{Accretion} &  Pulsations, & \multirow{2}{*}{Neutron star} \\
                         &                            & QPO          & \\
\midrule

\multirow{3}{*}{Hours} & Non-radial pulsations, & Line profile var. & \multirow{3}{*}{Be star} \\
                        & Rotation              & Multi-periodic    & \\
                        &                       & lightcurves       & \\
\midrule

Weeks to & Density perturbations,            & V/R var.          & \multirow{2}{*}{Be disk} \\
Months   &  \multirow{2}{*}{Orbital motion}  & Brightness var.   & \multirow{2}{*}{Binary} \\
         &                                   & X-ray outbursts   & \\
\midrule

\multirow{4}{*}{Years} & \multirow{2}{*}{Disk formation} & Line profile var. & \multirow{4}{*}{Be disk} \\
                       &                                  & Line strength var. & \\
                      & \multirow{2}{*}{and dissipation} & Brightness var. & \\
                      &                 & Color changes     &\\
\bottomrule
\end{tabularx}
%\end{adjustwidth}
\end{table}
\unskip

%--------------------------------------------------------------

\section{The Be~Star}

The origin of the Be phenomenon, that is, the~identification of the ultimate
mechanism that leads to the formation of a circumstellar disk around the equator
of a Be star is one of the major open questions in the field. Although~there is
general consensus that the disk formation is related to the high rotational
velocity of the Be star, there is no agreement on how close to critical velocity
(velocity at which centrifugal forces balance Newtonian gravity) they
rotate. Rotation rates (rotation velocity normalized by the star's critical
rotation speed) lie in the range 70--80\% \cite{porter03, rivinius13}, 90\%
\cite{fremat05}, or~even higher~\cite{townsend04}.  Several scenarios have been
put forward to explain these fast rotation rates~\cite{carciofi25}: (i) Be stars
are born as fast rotators; (ii) the high rotation speeds result through
accretion from a companion during an evolutionary phase as a binary system;
(iii) they become fast rotators as they evolve along the main sequence.
The~rotation rate is a key ingredient in understanding the formation of the
equatorial disk. The~higher the rotation rate, the~easier it is for gas to lift
up and launch material into a ballistic orbit. Still, some other mechanism must
be at play. The~primary trigger for episodic mass ejection into a disk is
non-radial pulsations (NRP). Constructive interference or beating between two or
more NRP modes may contribute to the sudden increases in pulsation amplitude and
lead to a mass outburst~\cite{baade17,baade18b}.

Observationally, NRPs manifest as short-term (hours) variability.
NRPs divide the stellar surface in regions with different temperatures and
velocity fields. If~the amplitude of the pulsations is large enough, the~temperature variations across the surface of the star can be detected
photometrically as differences in brightness~\cite{baade16,semaan18,balona21},
while the redistribution of flux over the absorption line profile as the star
rotates creates moving patterns of peaks and troughs that can be detected
spectroscopically~\cite{baade84,rivinius03,zima06}.

NRPs are a very common, if~not ubiquitous, feature intrinsic of BeXBs~\cite{reig22}. As~in classical Be stars, BeXBs display frequency groups,
long-term irregular variability, isolated signals, and~light outbursts. This
type of variability in BeXBs occurs mainly on timescales shorter than $\sim$12
h. The~peaks of the period distribution correspond to modulations of \mbox{5--10 h},
but there is also a significant contribution on timescales $\sim$1~day.

\begin{figure}[H]
%\centering
\includegraphics[width=9.0 cm]{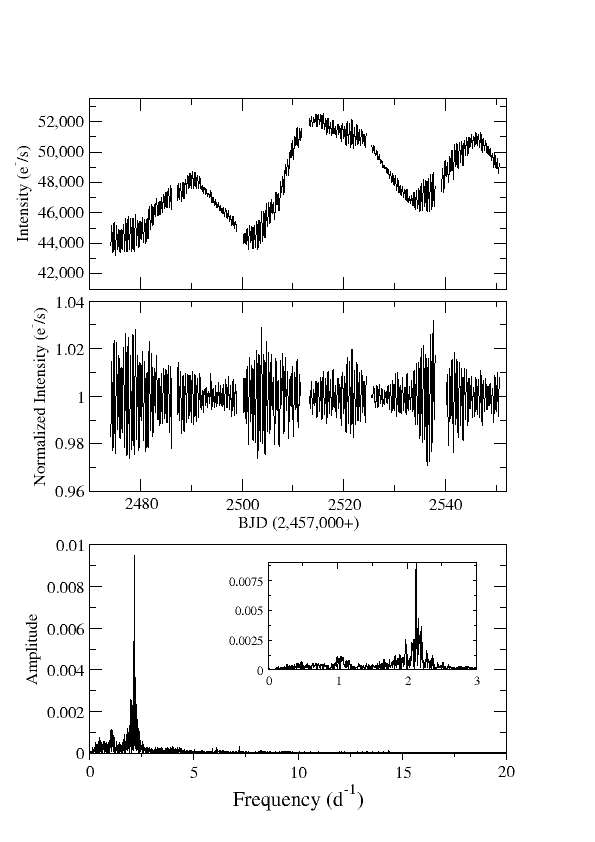}
\caption{(\textbf{Top panel}): 
TESS light curve of 1A 0535+26. (\textbf{Middle panel}): normalized light curve. (\textbf{Bottom panel}): periodogram. Note the largest amplitude of the oscillation during the brightening parts of the light curve.  Adapted from~\cite{reig22}. \label{nrp}}
\end{figure}

Figure~\ref{nrp} shows the {\bf \it Transiting Exoplanet Survey Satellite}
(TESS) light curve of 1A 0535+262 (top panel), the~normalized light curve
(middle panel), and~the corresponding periodogram (bottom panel). The~frequency
analysis reveals the characteristic presence of two frequency groups, centered
around 1 and 2 d$^{-1}$. Notably, the~oscillation amplitude increases after each
local minimum and then gradually decreases along the ascending branch toward the
brightness maxima. Following each maximum, the~amplitude reaches its lowest
values. This behavior is consistent with the general trend observed in classical
Be stars, in~which an increase in the mean stellar brightness is associated with
the enhancement of the amplitude of one or more pulsation frequency groups,
which in turn, are identified with mass ejection
episodes~\cite{labadie-bartz22}.

%\unskip   
%--------------------------------------------------------------

An alternative mechanism (or in addition to NRPs) that may eject material from
the photosphere to form the disk is magnetic reconnection associated with
localized magnetic fields generated by convection~\cite{balona21}.

\section{The Decretion~Disk}

As explained in the previous section, the~decretion disk forms when material is
ejected from the stellar photosphere due to mechanisms such as fast rotation,
non-radial pulsations, or~magnetic reconnection. What happens to the material
after it is ejected remains an open question, although~significant theoretical
progress has been made in recent years~\cite{panoglou16,cyr17,rubio25}.
Observationally, changes related to disk evolution are evident in the optical/IR
and X-ray bands.  Photometrically, the~contribution of the disk to the overall
brightness increases at longer wavelengths, producing an infrared excess and
reddening colors as the disk develops~\cite{reig15}. Spectroscopically, hydrogen
emission lines, particularly H$\alpha$,  show significant variations in profile
and strength, reflecting structural changes in the disk~\cite{reig16}.
Polarimetrically, Thomson scattering produces continuum polarization that traces
disk's density and orientation changes~\cite{reig26}. Disk size and density
determine how much material the neutron star can accrete, thereby regulating the
occurrence of X-ray~outbursts.

\subsection{Disk~Truncation}

Disks in BeXBs are smaller and denser than those around isolated Be stars. This
is because the neutron star's tidal forces truncate the disk.  Both observations~\cite{reig97,reig16} and theory~\cite{okazaki01,okazaki02,brown19} support disk
truncation in BeXBs. The~truncation radius, defined as the distance where tidal
and viscous forces balance, depends on disk parameters such as kinematic
viscosity (which measures internal friction) and aspect ratio (the disk's
thickness relative to its radius), as~well as binary parameters including
orbital period and eccentricity~\cite{okazaki01,panoglou16,cyr17,martin24}.

\begin{figure}[H]
\begin{tabular}{l@{\hspace{2mm}}l}
\includegraphics[width=6.5 cm]{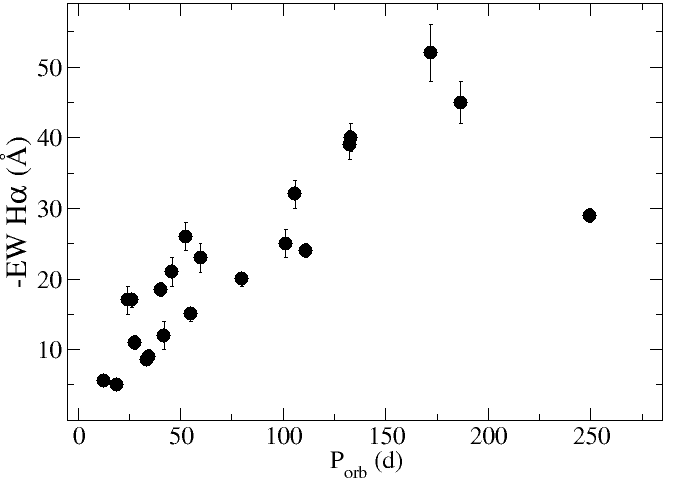} & 
\includegraphics[width=6.5 cm]{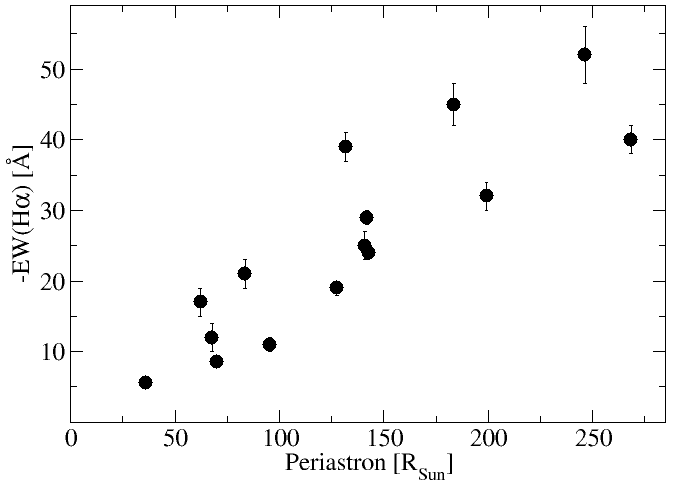} \\
\end{tabular}
\caption{H$\alpha$ equivalent width as a function of the orbital period (\textbf{left}) and periastron distance (\textbf{right}). 
\label{truncation}}
\end{figure}

The first observational evidence for truncation came from the correlation between the equivalent width of the H$\alpha$ line and the orbital period~\cite{reig97,reig16}. Systems with narrow orbits experience stronger tidal
forces from the neutron star. As~a result, their disks cannot maintain a stable
configuration over long timescales. In~contrast, systems where the neutron star
orbits farther away (long orbital periods) allow the disk to develop more
freely. Because~the \ha\ equivalent width measures the disk's extent~\cite{tycner05,grundstrom06}, this correlation implies that wider-orbit systems
develop larger and more stable disks. Figure~\ref{truncation} shows an updated
version of the $EW(H\alpha)$-$P_{\rm orb}$ diagram for Galactic BeXBs. The~system
with the longest orbital period shown in Figure~\ref{truncation} (left panel) that
lies outside the general trend is GRO J1008--57, which has the most eccentric
orbit ($e = 0.7$). Although~orbital period is easier to determine
observationally, the~truncation radius is expected to be determined by the
periastron distance, which for highly eccentric orbits would be comparable to
systems with shorter orbital periods. The~right panel in Figure~\ref{truncation}
shows the relationship between \ha\ equivalent width and periastron distance.
GRO J1008--57 with $\sim$142~$\rsun$ lies confortable in the middle of the
correlation.

\begin{figure}[H]
\begin{tabular}{c@{\hspace{2mm}}c}
\includegraphics[width=6.5 cm]{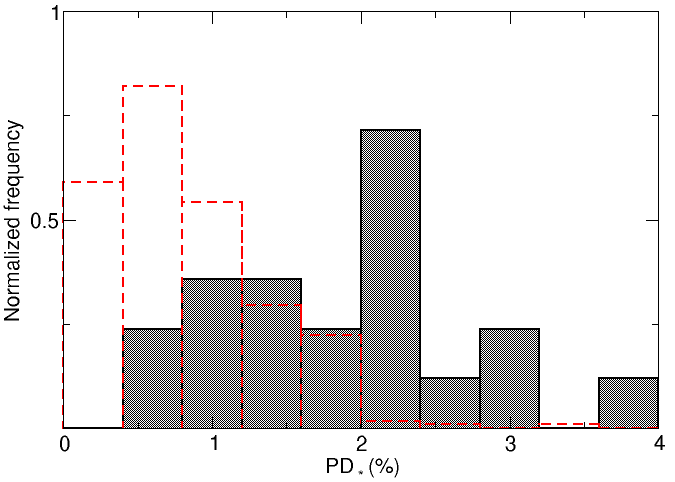} & 
\includegraphics[width=6.5 cm]{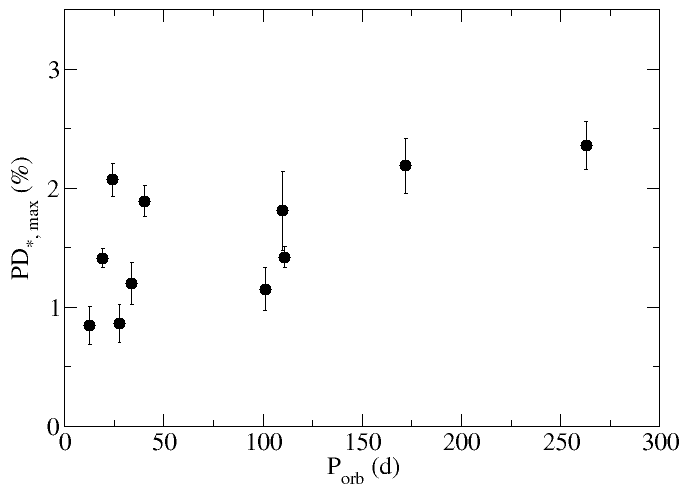} \\
\end{tabular}
\caption{{(\textbf{Left}):} Normalized distribution of the intrinsic polarization degree of classical 
Be stars (red dashed line) and BeXBs (black shaded rectangles). {(\textbf{Right})}: Intrinsic polarization degree as
a function of the orbital period. Adapted from~\cite{reig26}. \label{PDmax-hist}}
\end{figure}

Polarization studies further support the idea that BeXBs have denser, truncated
disks compared to classical Be stars. On~average, BeXBs exhibit higher intrinsic
polarization than isolated Be stars~\cite{reig26}. Since polarization degree
reflects the number of scattering electrons, models predict that polarization
increases with disk density due to more electrons available for Thomson
scattering~\cite{halonen13}. Figure~\ref{PDmax-hist} compares the normalized
distribution of intrinsic polarization degrees for classical Be stars (spectral
types O9--B2) and BeXBs. The normalized histogram is computed so that
the total area under the histogram equals 1, i.e., it represents an estimate of
a probability density function: $p(x_i)\approx\frac{n_i}{N \Delta x}$, where
$n_i$ is the number of measurements in bin $i$, $\Delta$x is the bin width, and~$N$ the total number of measurements. Here $N=314$ for Be stars, $N=11$ for
BeXB, and $\Delta$x~=~0.4. The~data are from~\cite{yudin01} (Be stars) and~\cite{reig26} (BeXBs). Be stars generally show lower polarization; notably, only
five stars in the~\cite{yudin01} sample had polarization degrees above 2\%, all
within the B0--B2 spectral range. In~contrast, about half of the BeXB sample
displayed maximum intrinsic polarization degrees above 2\%. Moreover,
Figure~\ref{PDmax-hist} shows that systems with longer orbital periods tend to
have higher polarization fractions, reinforcing the idea that the neutron star
limits the disk's free~expansion.

%--------------------------------------------------------------
\unskip   
%--------------------------------------------------------------
\vspace{-6pt}

%--------------------------------------------------------------
%--------------------------------------------------------------

\subsection{Disk Formation and Dissipation~Cycles}

The disk evolves on timescales of years, undergoing formation and dissipation
phases that control both long-term optical/infrared variability and X-ray
activity through mass supply for~accretion.  

\begin{figure}[H]
\includegraphics[width=.99\linewidth]{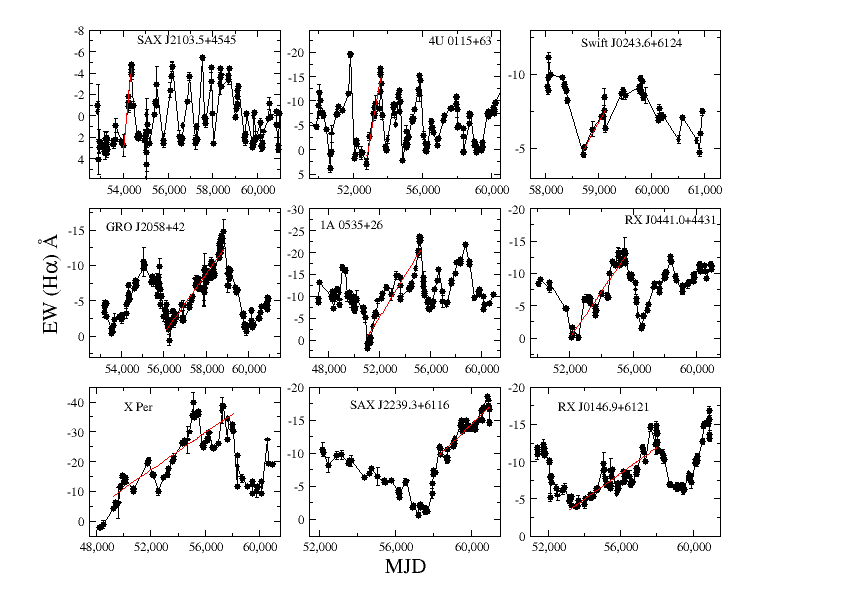}
\caption{Evolution 
%MDPI: Please change the hyphen (-) into a minus sign (−, “U+2212”) in the figure, e.g., “-1” should be “−1”.
 of the \ew. The  red lines represent a linear fit to the
data points that correspond to the formation of the circumstellar disk after a low-optical 
state. \label{EWHa}}
\end{figure}

Figure~\ref{EWHa} displays the evolution of the \ha\ equivalent width (\ew) for
several BeXBs, ordered in increasing value of the orbital
period. This plot has an enormous diagnostic power as it covers the formation,
growth, and~dissipation of the decretion disk. The~optical emission lines in Be
stars, most prominently the H$\alpha$ line, are formed through recombination
within the dense, ionized gas of the circumstellar decretion disk, where intense
ultraviolet radiation from the central B star ionizes hydrogen atoms, and~the
subsequent recombination of electrons with protons produces line emission.
Consequently, \ew\ serves as a robust proxy for the
presence or absence of the disk. Moreover, the~size of the disk is directly
related with the strength of the \ha\ line. This has been made evident through
interferometric observations that have resolved the disk and enabled the
measurement of the angular dimensions of Be star disks~\cite{quirrenbach97,
tycner05, grundstrom06}. By~convention, \ew\ is negative for emission lines and
positive for absorption~lines.

%--------------------------------------------------------------
%\unskip   
%--------------------------------------------------------------

The timescale for a whole cycle is given in Table~\ref{cyc}. There is a trend
for longer formation-dissipation timescales in systems with longer orbital
periods. A~correlation between the orbital period and the so-called superorbital
periods has been reported in many different types of X-ray binaries, although~the physical mechanism behind the superorbital modulation differs~\cite{weng24}.
In BeXBs, the~superorbital modulation is driven by long-term evolution of the Be
star's decretion disk, making this variability quasi-periodic rather than
strictly~coherent.

%---------------------------------------------------------------------------
\begin{table}[H]
%\centering
\caption{Decretion disk formation and dissipation timescales in~BeXBs.\label{cyc}}
%\begin{tabular}{lcc}
\begin{tabularx} {\textwidth} {lCC}
\toprule
\textbf{Source}              &\boldmath{$P_{\rm orb}$}      &\boldmath{$P_{\rm cycle}$}  \\
                    &\textbf{(d)}                &\textbf{(d)}           \\
\midrule
SAX J2103.5+4545    &12.7               &600    \\
4U 0115+63          &24.3               &1000   \\
IGR J06074+2205     &80~\textsuperscript{$\dag$}  &800    \\
GRO J2058+42        &55                 &3200   \\
1A 0535+262         &110                &1700   \\
RX J0441.0+4431     &150                &3500   \\
RX J0146.9+6121     &330~$^{\dag\dag}$   &3000   \\
\bottomrule
\end{tabularx}

\noindent\footnotesize{\textsuperscript{$\dag$} $P_{\rm  orb}=80/n$, where $n=2,3,\dots$; 
		\textsuperscript{$\dag\dag$} Tentative.}
\end{table}
%\unskip
%---------------------------------------------------------------------------

\subsubsection{Disk-Loss~Episodes}

Although the disk is central to understanding the observed variability of BeXBs
across multiple wavelengths, disk-loss events provide a unique opportunity to
disentangle the intrinsic stellar emission from disk-related contributions,
enabling a more reliable determination of the Be star's fundamental parameters.
Similarly, the~total reddening and polarization measured during a disk-loss
episode should correspond entirely to the interstellar medium. Disk-loss
episodes are best identified from spectroscopic observations when the \ha\ line
profile reverts from emission into absorption, or~equivalently, when the \ew\
changes from negative to positive in Figure~\ref{EWHa}. Therefore, disk-loss
episodes can be easily identified in Figure~\ref{EWHa} by looking at epochs when
\ew\ $> 0$.

\subsubsection{Disk~Growth}

\begin{figure}[H]
%\centering
\includegraphics[width=10.0 cm]{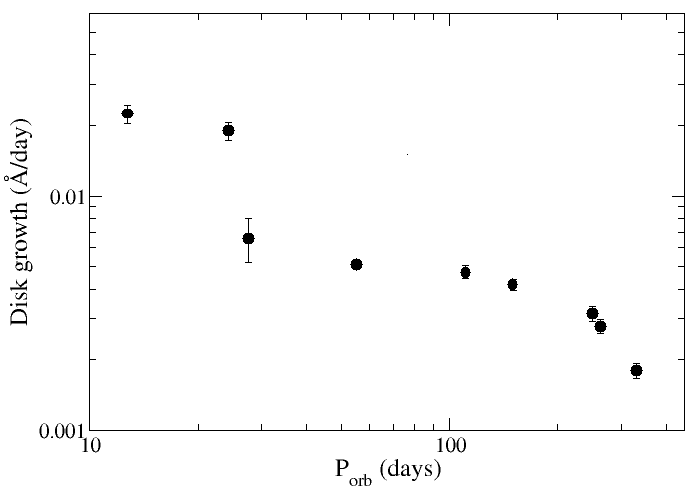} 
\caption{Disk growth rate after a disk-loss episode as a function of orbital~period. \label{growth}}
\end{figure}

Figure~\ref{EWHa} can also be used to estimate the disk growth rate, 
$d(EW(H\alpha))/dt = \dot{EW}$(\ha),
following an optically faint (low) state: systems with shorter orbital periods
exhibit faster growth rates. This correlation is further illustrated in
Figure~\ref{growth}, which plots the growth rate as a function of orbital period.
Since the mass ejection mechanism originates in the Be star and all BeXBs host
spectrally similar companions (with optical counterparts ranging from O9 to B2),
the observed correlation cannot be attributed to differences in mass ejection
mechanisms, which would lack a plausible physical explanation. Instead, the~trend is naturally explained by the tidal torque exerted by the neutron star,
which drives an accumulation effect wherein material builds up in the inner
disk. In~wider-orbit systems, the~neutron star's weaker tidal influence allows
the disk to extend farther, resulting in a more gradual radial decline in
surface density. By~contrast, in~close-orbit systems, the~neutron star's
proximity truncates the disk more severely, producing a steeper density~gradient.

%--------------------------------------------------------------
%\unskip   
%--------------------------------------------------------------

%\vspace{-6pt}

\begin{figure}[H]
%\centering
\includegraphics[width=.8\linewidth]{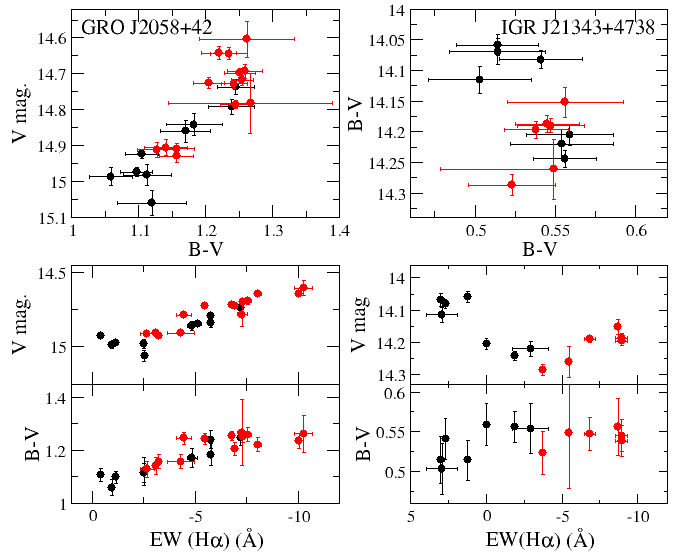} 
\caption{Evolution of the $V$ magnitude and ($B$-$V$) color during the growth of the disk after a low-optical state. Different color data points refer to two different epochs of disk~growth: black points correspond to observations in the interval 2005--2009 (GRO J2058+42) and 2009--2014 (IGR J21343+4738); red points correspond to the interval 2013--2019 (GRO J2058+42) and 2017--2020 (IGR J21343+4738). \label{phot}}
\end{figure}

Two types of correlations between photometric magnitudes and color indices are
observed during the formation and growth of the circumstellar disk, as~traced,
for example, by~increasing \ha\ emission strength. The~most common behaviour is
a positive correlation, in~which the brightness of the star (e.g., in~the $V$
band) increases simultaneously with the color index (e.g., {$B$-$V$}). In~other
words, as~the disk reforms and grows, the~system becomes both brighter and
redder. A~smaller number of stars exhibit a negative correlation. In~these
systems, stronger \ha\ emission is associated with a decrease in optical
brightness. The~fading observed in the $V$ band is again accompanied by a
reddening of the {($B$-$V$)} color index. These correlations are generally
interpreted as a geometrical effect~\cite{harmanec83}. Stars observed at very
high inclination angles tend to display the inverse (negative) correlation,
whereas systems viewed at inclination angles below a critical value, $i < i_{\rm
crit}$, show the positive correlation. In~nearly equator-on systems, the~inner
regions of the circumstellar disk partially obscure the stellar photosphere. At~the same time, the~small projected area of the disk on the plane of the sky
minimizes its contribution to the observed continuum emission. As~a result, the~system becomes fainter as the disk grows. In~contrast, for~low- and
intermediate-inclination systems, the~disk contributes additional continuum
emission and effectively increases the apparent radiating area of the
star-disk system. Consequently, disk growth is accompanied by an overall
increase in brightness. Although~the value of the critical inclination angle
remains uncertain, available observations suggest a representative value of
$i_{\rm crit} \approx 75^{\circ}$.
This interpretation is supported by the
larger number of systems that show a positive correlation than negative, the~fact that the same star always show the same type of correlation, and~also the
fact that when spectra are available, negative correlations are related to stars
that show shell spectra (a shell line is a spectral line that shows a
narrow absorption component superimposed on the broader stellar emission profile
and occurs when the Be disk is viewed at high inclination~\cite{hanuschik95}). Figure~\ref{phot}
shows two examples of the evolution of brightness and color as the disk
develops. GRO J2058+42 follows the more general case of increasing brightness
and reddening as the disk grows (i.e., as~\ew\ increases), corresponding to a
low- or intermediate-inclination system. On~the other hand, IGR J21343+4738
shows an overall drop in brightness as the disk grows, expected from a
high-inclination~system.

%--------------------------------------------------------------
%--------------------------------------------------------------

\subsubsection{Disk Formation and Dissipation: Predicting the Orbital Period of the~System}

If the observations encompass multiple cycles of disk formation and dissipation,
the evolution of the \ew\ could, in~principle, be used to estimate the orbital
period of systems lacking an orbital solution. This estimation is based on the
information presented in Figure~\ref{EWHa} and the relationships involving the
orbital period, specifically the $EW_{\rm max}$(\ha)--$P_{\rm orb}$ diagram and
the $\dot{EW}$(\ha)--$P_{\rm orb}$ relation.   The~results derived from these
diagrams can then be cross-checked for consistency using the less constrained 
$P_{\rm cyc}$--$P{\rm orb}$ diagram and the $PD_{\rm max}$--$P_{\rm orb}$
diagram. 

A notable feature of the data in Figure~\ref{EWHa} is the increasing smoothness of
the curves with longer orbital periods. Systems with shorter orbital periods
show many more disk formation and dissipation cycles, resulting in more ``choppy''
curves, whereas systems with longer orbital periods exhibit smoother
evolutionary~trends.

\begin{figure}[H]
%\centering
\includegraphics[width=.75\linewidth]{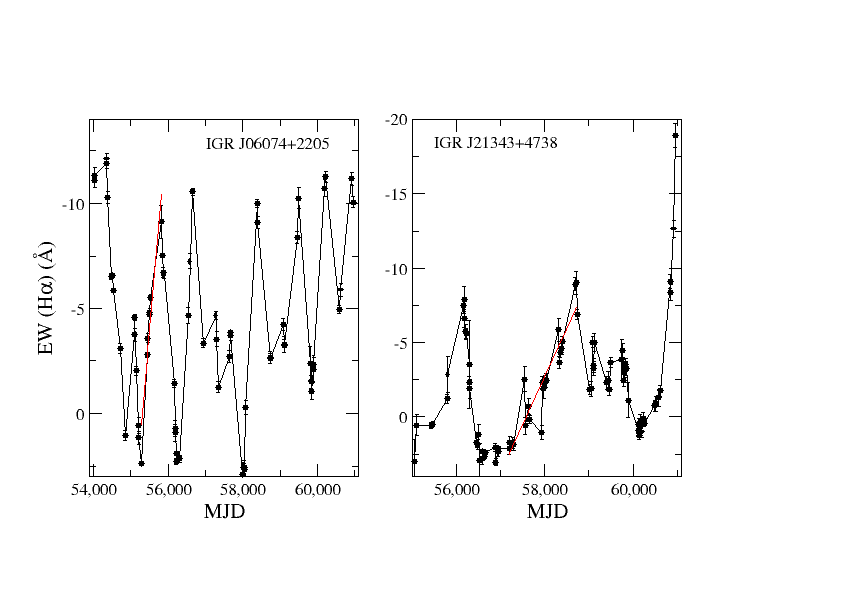} 
\caption{Evolution of the EW(\ha) for IGR J06074+2205 and  IGR J21343+4738. The  red lines represent a linear fit to the
data points that correspond to the formation of the circumstellar disk after a low-optical 
state.
\label{twocases}}
\end{figure}

To illustrate this method, we consider two cases: IGR J06074+2205 and IGR
J21343+4738. An~orbital period of 80 days has been proposed for IGR J06074+2205
based on the occurrence of relatively weak type I X-ray outbursts. However,
submultiples of this value could also be plausible, that is, 80/$n$, where
$n=2,3,\dots $~\cite{chhotaray24}. The~orbital period of IGR J21343+4738 remains
unknown. Figure~\ref{twocases} shows the evolution of \ew\ for these two
sources. From~this figure we can extract the following information: 

\begin{itemize}
    
    \item IGR J06074+2205: $EW_ {\rm max}$(\ha)~=~$-12.5$ \AA,
$\dot{EW}$(\ha)~$\approx0.017\pm0.003$ \AA/day, $P_{\rm cyc}\approx800$~days.
Also we know that $PD_{\rm max}=1.7$\%

    \item IGR J21343+4738: $EW_ {\rm max}$(\ha)~=~$-20$ \AA,
$\dot{EW}$(\ha)~$\approx0.007\pm0.001$ \AA/day, $P_{\rm cyc}\approx2400$~days.
Also, we know that $PD_{\rm max}=0.8$\% 

\end{itemize}

\begin{figure}[H]
%\centering
\includegraphics[width=.75\linewidth]{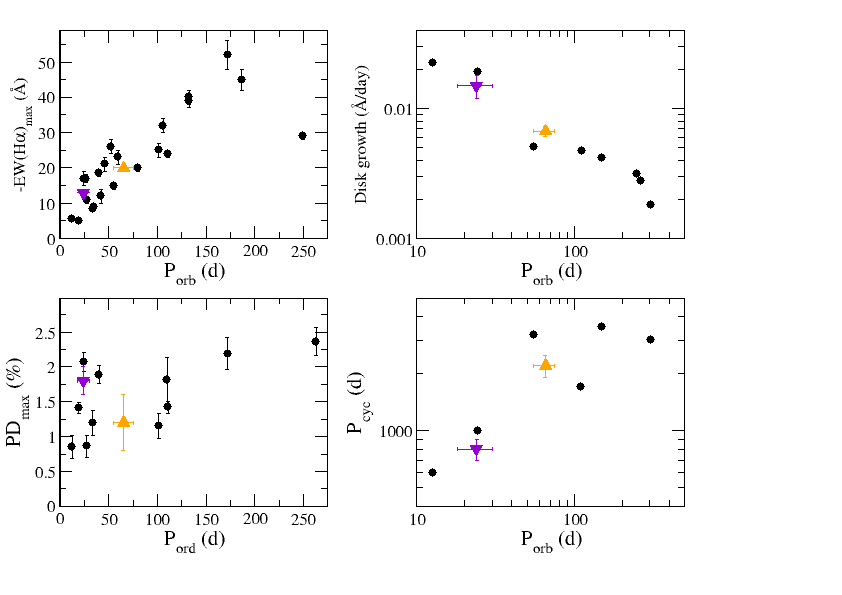} 
\caption{Various observable as a function of orbital period. The~colored points correspond to 
IGR J06074+2205 (violet) and IGR J21343+4738 (orange). \label{prediction}}
\end{figure}

Using $EW_ {\rm max}$(\ha) and $\dot{EW}$(\ha) and the best-fit in
Figures~\ref{truncation} and~\ref{growth}, we estimate the orbital period of IGR
J06074+2205 to be 18--30 days and predict that the orbital period of IGR
J21343+4738 should fall in the range 58--72 days. Figure~\ref{prediction} shows
the position of these two sources in the various~diagrams.

%--------------------------------------------------------------
%--------------------------------------------------------------
\vspace{-6pt}
%--------------------------------------------------------------
\unskip   
%--------------------------------------------------------------

\section{The Neutron~Star}

BeXBs host a strongly magnetized neutron star with magnetic fields in the order
of $10^{12}$--$10^{13}$ G. Information about the magnetic field strength
is obtained from the detection of cyclotron resonant scattering features (CRSF)
\cite{staubert19}. The~magnetic field is also responsible for the X-ray
pulsations.  Pulsations result from the misalignment of the neutron star's
rotational axis and the magnetic axis~\cite{rappaport77}.

\subsection{X-Ray Periodic, Quasi-periodic, and~Aperiodic~Variability}

Transient BeXBs exhibit strong X-ray variability. Timescales attributed to the
accretion process range from a fraction of a second to hours, resulting
in pulsations, quasi-periodic oscillations (QPOs) in the millihertz
range, and~band-limited noise in the power spectrum~\cite{revnivtsev09,mushtukov19,yang26}.

Periodic variability arises from the rotation of the neutron star and is
observed as coherent X-ray pulsations in the light curves. The~accretion flow is
funneled by the strong magnetic field onto the magnetic polar caps, where it
impacts the stellar surface and forms localized hot spots. In~these regions, the~kinetic energy of the infalling material is efficiently converted into X-ray
radiation through shock heating and subsequent radiative processes. If~the
magnetic axis is inclined relative to the rotation axis, the~hot spots sweep
across the observer's line of sight as the star spins, producing the observed
pulsations. The~corresponding spin periods typically range from about 1 to 1000
s. Variations in the spin period allow estimation of the magnetic field
strength and can be used to derive orbital solutions. Short-term
changes in pulse period, measured as delays in pulse arrival times caused
by the Doppler effect from orbital motion, help determine the orbital
parameters of the binary system~\cite{raichur10}. Long-term changes in the
pulse period are linked to the exchange
of angular momentum between the accretion flow and the neutron star. These
secular spin-up or spin-down trends provide valuable constraints on the neutron
star's magnetic field strength and on the physical properties and geometry of
the accretion flow~\cite{malacaria20}.

QPOs are associated with inhomogeneities in the accretion flow or plasma instabilities
generated around the magnetospheric boundary~\cite{yang26}. Two models have been
put forward to explain QPOs in accreting pulsars: the Magnetospheric
Beat-Frequency Model (BFM)~\cite{alpar85} and the Keplerian Fequency Model (KFM)~\cite{vanderklis87}. The~difference between the two is that the BFM proposes
that the QPO frequency is the beat between the neutron star's spin frequency and
the Keplerian frequency of matter at the inner edge of the accretion disk, while
the KFM suggests that the QPO frequency is directly equal to the Keplerian
orbital frequency at the inner disk. Thus, in~the BFM $\nu_{\rm QPO}=\nu_{\rm
K}-\nu_{\rm NS}$, while in the KFM $\nu_{\rm QPO}=\nu_{\rm K}$. Assuming the
radius of the inner accretion disk scales by the accretion rate (and hence the
X-ray luminosity), the~Keplerian frequency of the inner accretion disk, and~thereby the QPO frequency, is expected to increase with luminosity in both KFM
and BFM, as~observed~\cite{ma22}. A~third model, the~magnetic disk precession model~\cite{shirakawa02}, attributes mHz QPOs to warping/precession modes induced by
magnetic torques near the inner edge of the accretion disk. None of the three
models can explain the origin of QPOs in all the systems. For~example, the~identification of the QPO frequency with the Keplerian frequency associated with
the disk magnetosphere boundary in the KFM imposes the condition $\nu_{\rm NS} <
\nu_{\rm QPO}$, which is not always~met.

\begin{figure}[H]
%\centering

\begin{adjustwidth}{-\extralength}{0cm}
\centering %% If there is a figure in wide page, please release command \centering, for Table, ``\textwidth" should be ``\fulllength"
\begin{tabular}{c@{\hspace{5mm}}c}
\includegraphics[width=8 cm]{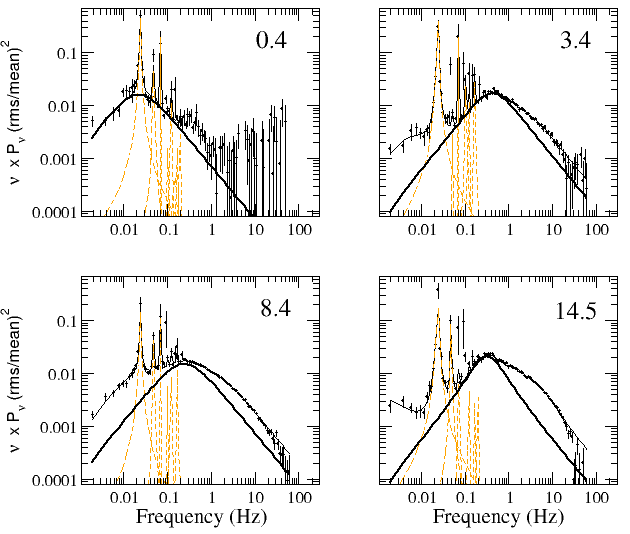}  &
\includegraphics[width=9 cm]{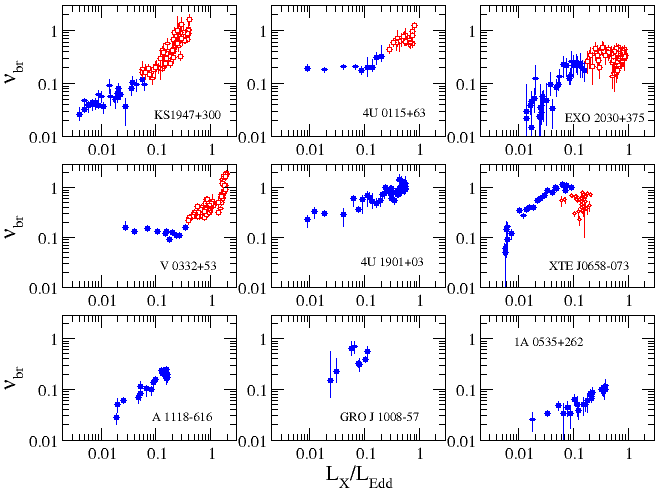}  \\
\end{tabular}
\end{adjustwidth}
\caption{{(\textbf{Left})}: Power spectrum of EXO 2030+375 at different X-ray luminosity (given 
at the top right of each panel in units of $10^{37}$ erg s$^{-1}$. Orange lines represent fits to the coherent pulsation and its harmonics, while the black line corresponds to the main broad-band noise component. Adapted from~\cite{reig08}. {(\textbf{Right})}: Break frequency versus X-ray luminosity for various
BeXBs. Blue and red circles correspond to the HB and DB, respectively {(see Section~\ref{accreg} for an explanation of these two states)}. \label{psd_exo}}
\end{figure}

Aperiodic variability appears as band-limited noise, whose power typically
decreases as the frequency increases. This noise is modeled by fitting the power
density spectra (PSD) with combinations of several Lorentzian~\cite{reig08,reig13} and/or (broken) power-law functions~\cite{doroshenko20,li24}. Accreting pulsars, hence BeXBs, generally display
breaks in their PDS.  Ref.~\cite{revnivtsev09} proposed that the break frequency is
directly related to the inner edge of the disk, specifically at the interface
between the accretion disk and the magnetosphere. Interestingly, at~low flux,
the noise power breaks occur near the pulse frequency, indicating corotation
between the neutron star and the accretion disk. As~the X-ray luminosity
increases, the~break frequency increases.  This correlation supports the
disk-magnetosphere origin of the frequency. Fluctuations of the X-ray flux are
largely caused by fluctuations of the mass accretion rate onto the neutron star
surface. An~increase in the mass accretion rate decreases the size of the
magnetosphere (and hence the inner radius of the disk), so that the
characteristic frequency at the inner edge of the disk/flow increases, and~the
inner parts of the accretion disks rotate much faster than the central object.  
This correlation between the break frequency and X-ray flux has been observed in
many sources~\cite{reig08,reig13,doroshenko20,reig22b,li24}. Some examples are
shown in Figure~\ref{psd_exo}.

%--------------------------------------------------------------
%--------------------------------------------------------------

\subsection{X-Ray~Outbursts}

The vast majority of BeXBs are hard X-ray transients that shine bright during
X-ray outbursts, which are traditionally divided into two types~\cite{motch91}:

\begin{itemize}

\item Type I (or normal) outbursts are relatively regular and often quasi-periodic
events, typically peaking near periastron passage of the neutron star. They
reach moderate X-ray luminosities, ($L_{\mathrm{X}} \lesssim 10^{37}$ erg
s$^{-1}$), and~usually last for a limited fraction of the orbital period
($\sim$0.2--0.5 ($P_{\mathrm{orb}}$)). These events are generally interpreted
as enhanced accretion episodes occurring when the neutron star interacts with
the outer regions of the Be star's circumstellar~disk.

\item Type II (or giant) outbursts are substantially brighter, with~luminosities
approaching or reaching the Eddington limit for a neutron star, ($L_{\mathrm{X}}
\sim 10^{38}$ erg s$^{-1}$). Unlike type I events, they show no preferred
orbital phase and may persist for a large fraction of an orbital cycle or even
several orbital periods. They are commonly associated with the accretion of a
significant portion of the Be disk. Current models
attribute these events to major structural changes in the circumstellar disc,
such as warping and the development of eccentricity in a highly misaligned disc,
which enable enhanced mass transfer onto the neutron star~\cite{okazaki13,martin14}.

\end{itemize}

\begin{figure}[H]
%\centering
\includegraphics[width=.8\linewidth]{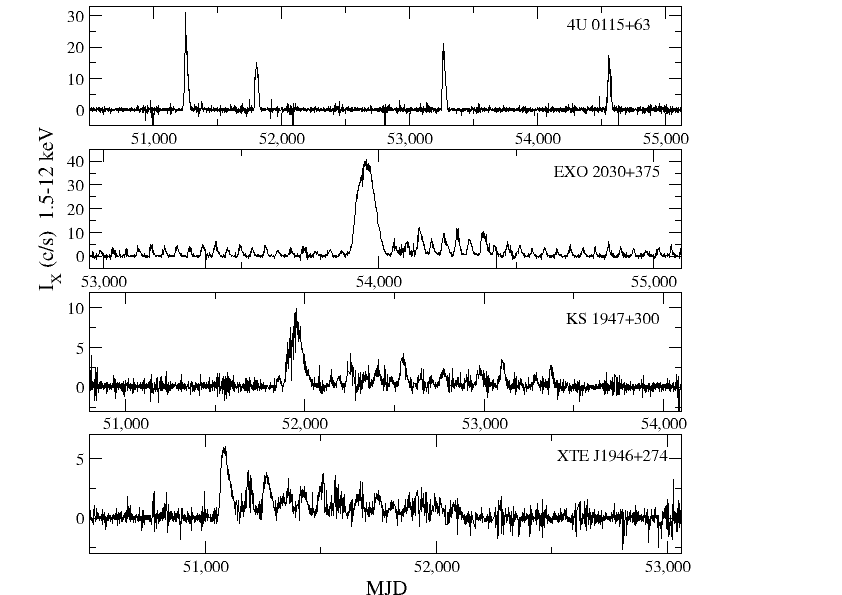} 
\caption{The two types of X-ray outbursts in BeXBs: low intensity, orbitally
modulated, and~shorter duration outbursts (type I) and giant and longer outbursts
(type II). In~this figure, 4U 0115+63 displays only type II outbursts, whereas
KS 1947+300 and XTE J1946+274 show a type II outburst followed by a series of
type I outbursts. EXO 2030+375 exhibits type I outbursts before and after a type
II outburst. The X-ray data were obtained with the RXTE/ASM All Sky Monitor. \label{xout}}
\end{figure}

Figure~\ref{xout} shows representative examples of these two type of X-ray
outbursts. It is not uncommon to observe a series of type I outbursts following a
major type II~event.

%--------------------------------------------------------------
\unskip   
%--------------------------------------------------------------

\subsection{Disk--Neutron Star~interaction}

Although it is commonly understood that the outbursts are caused by mass
transfer from the Be disk to the neutron star, the~detailed mechanism remains
uncertain. In~particular, the~connection between the Be disk and the resulting
X-ray activity is still debated. Type II outbursts are observed in systems with
both large and small disks. Furthermore, some systems with large disks exhibit
no X-ray activity~\cite{monagent17}.  Since the disk is truncated, significant
accretion can only occur if the disk becomes sufficiently asymmetric and dense
to overflow the truncation radius. The~prevailing hypothesis is that giant
outbursts happen when the neutron star captures a large amount of gas from a
warped, highly misaligned, and~eccentric Be disk. Models indicate that such
highly distorted disks lead to enhanced mass accretion when the neutron star
passes through the warped region~\cite{okazaki13,martin14}.

A mechanism proposed for the occurrence of type II outbursts in BeXBs is the
Kozai--Lidov (KL) mechanism~\cite{kozai62,lidov62}.  The~key idea in this
mechanism is that the product of the disk's inclination and eccentricity remains
constant. Thus, a~test particle that is initially on a circular orbit in a
misaligned disk undergoes a series of oscillations where the inclination and
eccentricity interchange periodically. The~eccentricity growth causes the disk
to overflow its Roche lobe and transfer material to the companion neutron star~\cite{martin14b,fu15}.

\begin{figure}[H]
%\centering
\includegraphics[width=.8\linewidth]{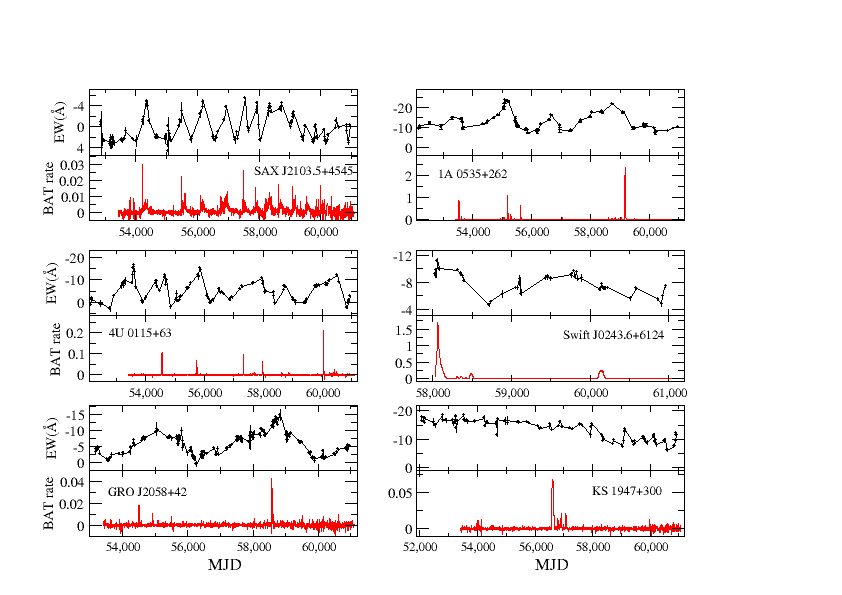} 
\caption{Optical (black) and X-ray (red) variability. After~major X-ray outbursts the Be
star's disk weakens. The X-ray data were obtained with the Swift/BAT Hard X-ray
Transient Monitor. \label{disk-NS}}
\end{figure}

Since the disk serves as the reservoir of material available for accretion, a~decrease in the size of the Be star disk should be observed following a major
outburst. Indeed, observations confirm the dissipation of the disk; however, the~extent of this loss, ranging from complete destruction to a minor
reduction, depends on the geometry and parameters of the orbit.
Figure~\ref{disk-NS} shows representative examples of the connection
between X-ray outbursts and the \ew. A~substantial decrease in \ew\ (hence in
the extent of the disk) is apparent after major X-ray~outbursts.

%--------------------------------------------------------------
\unskip   
%--------------------------------------------------------------

\subsection{Accretion~Regimes}
\label{accreg}

Most of the time, BeXBs are in a dormant or quiescent X-ray state and become
detectable only during outburst. Centrally important to their behavior, the~disk
acts as the primary reservoir of matter for accretion onto a compact object,
meaning its presence directly dictates X-ray variability and outbursts. If~the
disk dissipates completely, mass transfer to the companion star halts and X-ray
emission should cease. However, at~least three systems have shown X-ray emission
during disk-loss phases: 1A 0535+262~\cite{negueruela00}, IGR J21343+4738~\cite{reig14a}, and~SAX J2103.5+4545~\cite{reig14b}.  In~addition, many more
sources have been detected in a quiescent state at luminosities $L_X \simless
10^{34}$ erg s$^{-1}$ \cite{tsygankov17}. The~origin of the quiescent X-ray
emission remains unclear.  Two basic mechanisms have been proposed to explain
X-ray emission from accreting pulsars at very low luminosities: accretion and
cooling of the neutron star surface. Each mechanism implies different physical
conditions. Accretion may occur at the magnetosphere via a stellar wind or
through an accretion disk.  Cooling may originate from the entire neutron star
surface or from the polar caps. Timing and spectral analyses can help
distinguish between these scenarios. In~general, accretion leads to a power-law
dominated X-ray spectrum (due to Comptonization) and significant variability
(rms > 10\%), while cooling produces a thermal spectrum and a
white-noise-dominated power~spectrum.

During Type II outbursts, the~X-ray luminosity increases by several orders of
magnitude. This dramatic increase is associated with fundamental changes in the
accretion flow configuration. The~neutron star's strong magnetic field
disrupts the accretion flow at some distance from its surface and channels the
matter onto the magnetic polar caps, creating hot spots and an accretion column.
The conditions prevailing in the accretion column define two accretion regimes:
super- and sub-critical. These two regimes differ in the way the radiation
pressure of the emitting plasma is capable of decelerating the accretion flow.
In the super-critical regime, at~high luminosities, radiation pressure
dominates,  decelerating the accretion flow via photon interactions. 
A~radiation-dominated shock halts the flow at some distance above the neutron star
surface~\cite{davidson73,basko76,lyubarskii82}. In~the sub-critical regime,
radiation pressure is insufficient to stop the flow, which continues to the
surface, where it is decelerated by multiple Coulomb scatterings with thermal
electrons and nuclear collisions with atmospheric protons~\cite{burnard91,harding94}. The~luminosity marking the transition between these
regimes is known as the critical luminosity, $L_{\rm crit}$.

Observational evidence for this transition includes changes in pulse profiles
and correlations between cyclotron line energy and luminosity. As~X-ray
luminosity increases during an outburst, the~pulse profile typically changes
from a single to a double peak, interpreted as a change in accretion column
structure and the formation of a radiation shock. This corresponds to a change
from a pencil beam (photons escaping along the magnetic field axis) to a fan
beam (photons escaping sideways due to the shock)~\cite{wilson18,ji20}.
Because the cyclotron line energy is representative of the magnetic field
strength at the location where the line is formed, we expect its energy to
change with changes in the structure of the accretion column. A~bimodal behavior
has been observed. Some sources show a
positive correlation between cyclotron energy and luminosity (indicative of the
sub-critical regime), whereas others show an anticorrelation (indicative of the
super-critical regime)~\cite{becker12,mushtukov15}.

\begin{figure}[H]
%\centering
\begin{tabular}{ll}
\includegraphics[width=6.5 cm]{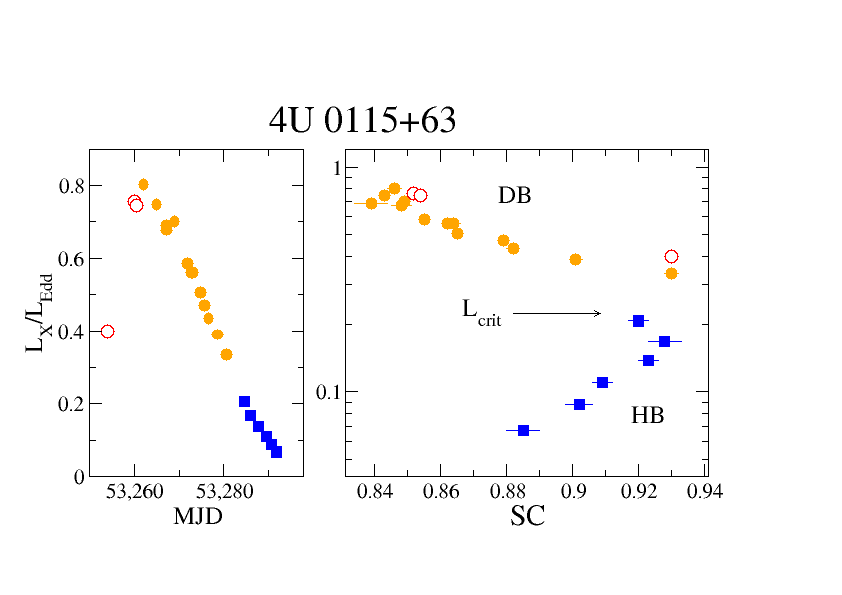}\hspace{1mm} &
\includegraphics[width=6.5 cm]{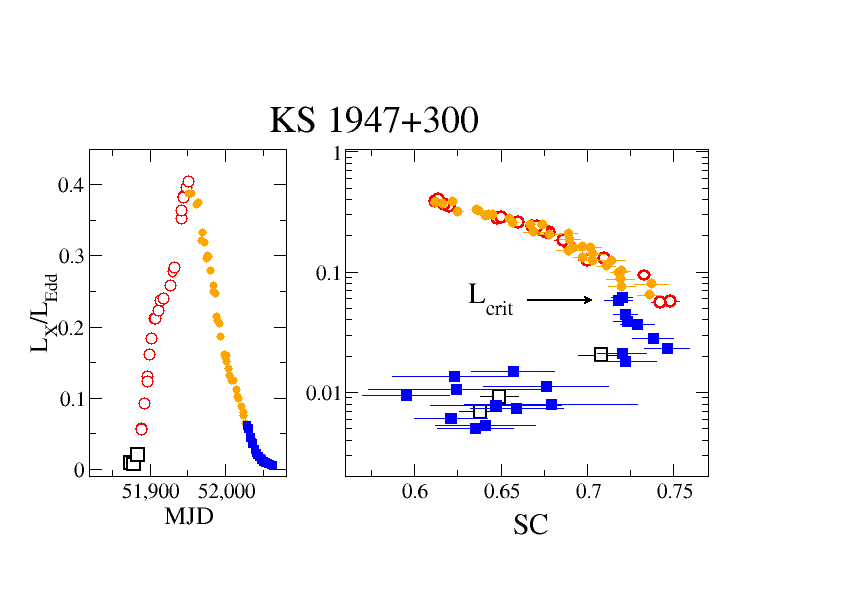}\hspace{1mm} \\[0.5mm] % small vertical gap
\includegraphics[width=6.5 cm]{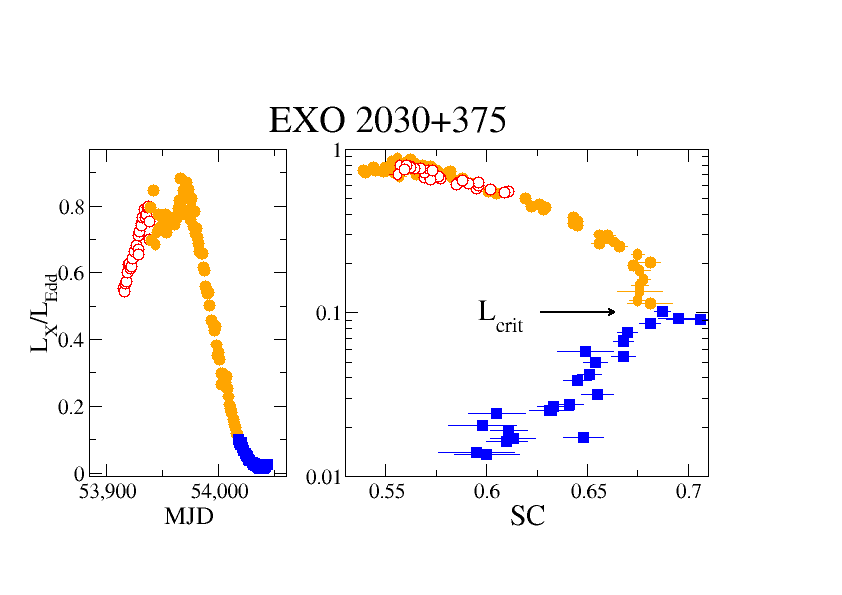}\hspace{1mm} & 
\includegraphics[width=6.5 cm]{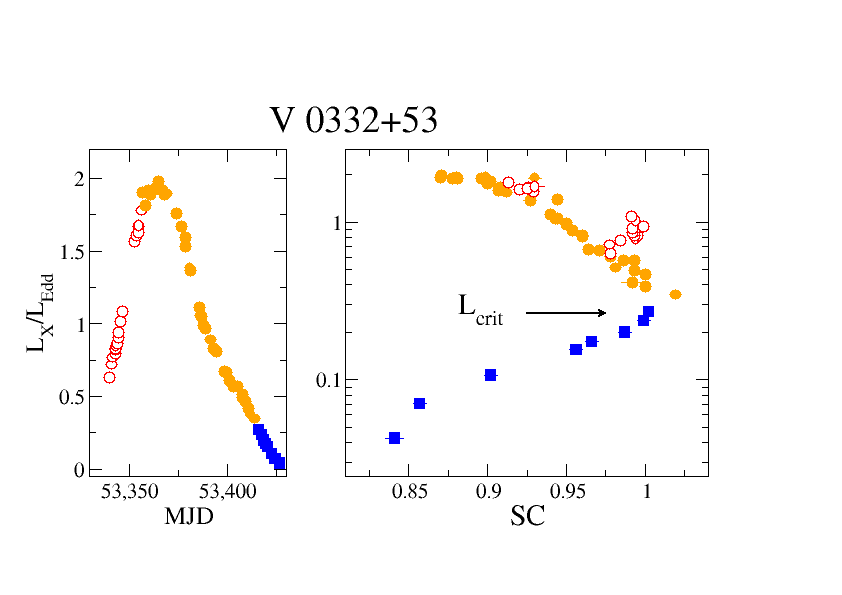}\hspace{1mm} \\
\end{tabular}
\caption{HID  for four BeXBs during a type II outburst. Different colors
represent different instances during the X-ray outbursts as indicated in the
light curve (smaller panels) of each source. The soft colour (SC) is
defined as the ratio 7--10 keV/4--7 keV. Typically, the~Eddington luminosity
for a neutron star is $1.7 \times 10^{38}$ erg s$^{-1}$. The data were obtained
with the RXTE/PCA instrument. \label{crit}}
\end{figure}

Further evidence for the transition between accretion regimes comes from
variability in the X-ray spectral continuum. The~hardness--intensity diagram
(HID) is a useful tool to describe the rich variability patterns during
outbursts. The~source traces different branches or states in this diagram as it
progresses through an outburst. Two main branches are observed in BeXBs: the
``horizontal'' branch (HB), corresponding to the low-luminosity sub-critical
state, and~the ``diagonal'' branch (DB), associated with the high-luminosity
super-critical state~\cite{reig13}. The~luminosity at which the source
transitions between these branches would naturally correspond to the critical
luminosity.  From~a theoretical perspective, $L_{\rm{crit}}$ depends on many
factors, including the neutron star's mass and radius, the~size of the hot
spots, magnetic field strength, temperature, photon energy, and polarization~\cite{basko76,becker12,mushtukov15}.
Generally, larger magnetic fields correspond to higher critical luminosities.
Since different systems have different magnetic fields, the~sub-critical to
super-critical transition must occur at different luminosities, consistent with
observations. Figure~\ref{crit} shows examples of HIDs illustrating this
behavior.
\vspace{-6pt}

%--------------------------------------------------------------
%--------------------------------------------------------------

\begin{figure}[H]
%\centering
\begin{tabular}{ll}
\includegraphics[width=6.5 cm]{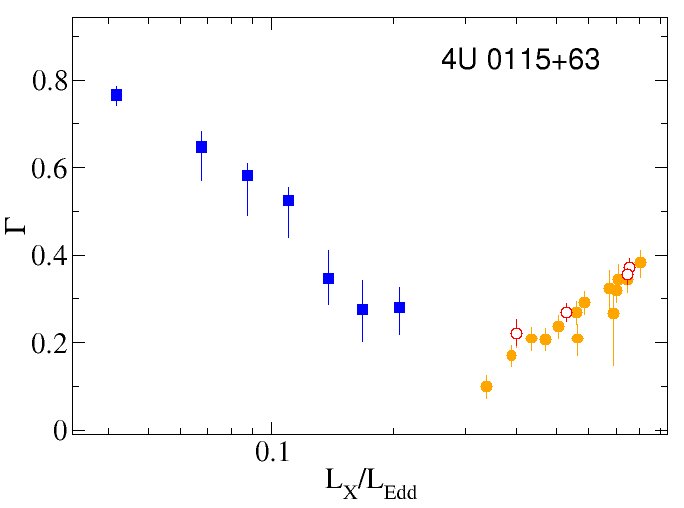}\hspace{1mm} &
\includegraphics[width=6.5 cm]{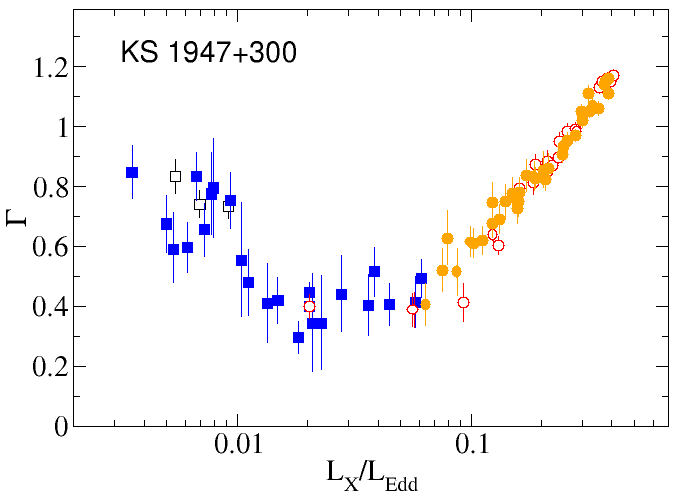}\hspace{1mm} \\[0.5mm] % small vertical gap
\includegraphics[width=6.5 cm]{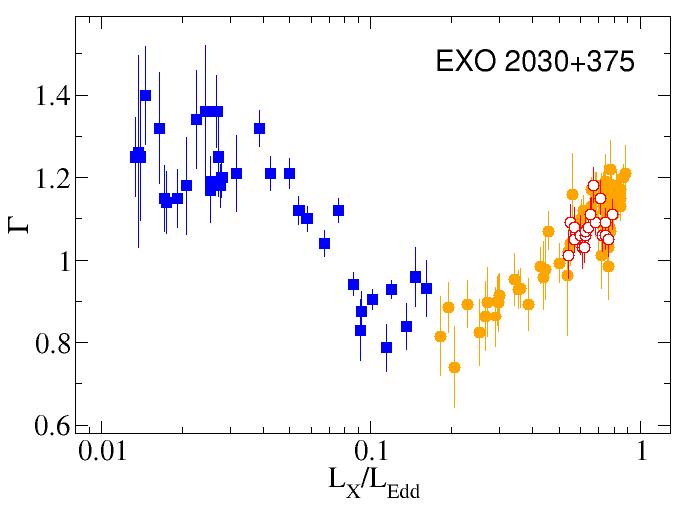}\hspace{1mm} & 
\includegraphics[width=6.5 cm]{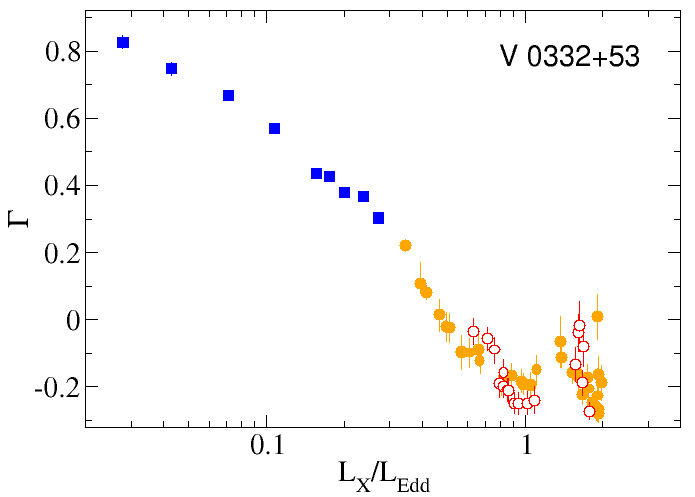}\hspace{1mm} \\
\end{tabular}
\caption{Photon index $\Gamma$ as a function of X-ray luminosity. Symbols as in
Figure~\ref{crit}.\label{gamma}}
\end{figure}

The typical energy spectral continuum of an accreting pulsar is well described
by an absorbed power-law and exponential cutoff. The~absorption at the lower
energy end of the spectrum is caused by interstellar gas through the
photoelectric effect. The~power-law and exponential cut off at the higher energy
of the spectrum is attributed to comptonization. Hardness ratios (HR) represent
the simplest way to characterize the X-ray spectral continuum. Chosing them
appropriately, the~HR is a measure of the slope of the power law. However, a~more accurate measure of the spectral continuun is provided by the photon index
(i.e., the~exponent of the power law). Similar to the HR, the~photon index
$\Gamma$ also exhibit a break when  plotted as a function of X-ray luminosity or
flux~\cite{reig13,serim22,mandal23}, indicating a major change in the
configuration of the accretion flow and/or the structure of the accretion
column; i.e.,~it marks the transition between states (Figure~\ref{gamma}).

%\vspace{-12pt}

%--------------------------------------------------------------
%--------------------------------------------------------------

\section{Conclusions}

Be/X-ray binaries constitute the largest and most diverse subgroup of neutron
star high-mass X-ray binaries, serving as invaluable laboratories for the study
of accretion physics, stellar evolution, and~disk dynamics. Their hallmark
feature, the~decretion disk formed around the Be star, plays a crucial role in
regulating the mass transfer to the neutron star and thereby dictates the rich
variability observed across multiple wavelengths and timescales. The~complex
interplay between the Be star's rapid rotation, non-radial pulsations, and~disk formation mechanisms underpins the long-term optical and X-ray variability
characteristic of these~systems.

Observational evidence supports the truncation of the Be star's disk by the
neutron star's tidal forces, which influences disk size, density, and~consequent X-ray outburst behavior. The~neutron star's strong magnetic
field governs the accretion flow geometry and pulsation characteristics, with~transitions between sub-critical and super-critical accretion regimes evident in
spectral and timing~properties.

The ongoing study of Be/X-ray binaries continues to reveal new complexities,
including persistent low-luminosity systems and variations in accretion regimes,
underscoring the need for coordinated multiwavelength observations. These
systems not only enrich our understanding of binary star evolution and
high-energy astrophysics but also provide unique testbeds for theoretical models
of disk dynamics, mass transfer, and~magnetically channeled~accretion.

In summary, Be/X-ray binaries are multifaceted astrophysical systems whose
detailed study sheds light on fundamental processes in stellar and compact
object astrophysics, and~sustained observational efforts promise further
advances in unraveling their intriguing~behaviors.

%%%%%%%%%%%%%%%%%%%%%%%%%%%%%%%%%%%%%%%%%%
\vspace{6pt}

%%%%%%%%%%%%%%%%%%%%%%%%%%%%%%%%%%%%%%%%%%

\funding{This research received no external~funding. }

%\institutionalreview{Not applicable.}

%\informedconsent{Not applicable.}

\dataavailability{Most of the data presented in this study have been published
and can be found on the corresponding publication. They are also available on request
from the corresponding author.} 

\acknowledgments{This research has made use of the SIMBAD database, operated at
CDS, Strasbourg, France. The optical data were obtained using the facilities at
the Skinakas Observatory, which is a joint research facility of the University
of Crete and the Foundation for Research and Technology \u2013 Hellas (FORTH).}

\conflictsofinterest{The author declares no conflicts of~interest.}

%\abbreviations{Abbreviations}{
%The following abbreviations are used in this manuscript:
%HMXB: High-mass X-ray binary\\
%BeXB: Be/X-ray binary\\
%}

\noindent 

%\appendixtitles{yes} % Leave argument "no" if all appendix headings stay EMPTY (then no dot is printed after "Appendix A"). If the appendix sections contain a heading then change the argument to "yes".
%\appendixstart
%\appendix
%\section[Galactic BeXBs]{}

%%%%%%%%%%%%%%%%%%%%%%%%%%%%%%%%%%%%%%%%%%
%\isPreprints{}{% This command is only used for ``preprints''.
\begin{adjustwidth}{-\extralength}{0cm}

\reftitle{\highlighting{References} %MDPI: Please note that we spell out the names of the journal name in the reference list. We have completed this revision throughout the manuscript; please confirm all reference journal names.
}

% Please provide either the correct journal abbreviation (e.g., according to the “List of Title Word Abbreviations” http://www.issn.org/services/online-services/access-to-the-ltwa/) or the full name of the journal.

% \bibliography{your_external_BibTeX_file}

%=====================================
% References, variant B: internal bibliography
%=====================================

% ACS format
%\isAPAandChicago{}{%

\PublishersNote{}
%\isPreprints{}{% This command is only used for ``preprints''.
\end{adjustwidth}
\end{document}